\documentclass[9pt,twocolumn,twoside]{pnas-new}

\usepackage[normalem]{ulem}

\templatetype{pnasresearcharticle} 

\makeatletter
\@ifpackageloaded{stix} 
{
}{                      
	\usepackage{mathrsfs}
	\usepackage{txfonts}              
}

\ifx\bm\undefined
  \newcommand{\bm}[1]{{\boldsymbol {\mathrm #1}}} 
\else
  \renewcommand{\bm}[1]{{\boldsymbol {\mathrm #1}}} 
\fi
\makeatother

\newcommand{\f}[2]{\frac{#1}{#2}}
\newcommand{\mr}[1]{\mathrm{#1}}

\newcommand{\para}{{||}}

\makeatletter
\newcommand{\vast}{\bBigg@{4}}
\newcommand{\Vast}{\bBigg@{5}}

\newcommand\agk{{\tt AstroGK}}

\newcommand{\krhoi}{k_\perp \rho_\mathrm{i}}

\newcommand{\add}[1]{{#1}}
\makeatother

\DeclareMathAlphabet{\mathpzc}{OT1}{pzc}{m}{it}

\newcommand{\rmc}{\mathrm{c}}
\newcommand{\rmd}{\mathrm{d}}
\newcommand{\rme}{\mathrm{e}}

\newcommand{\rmi}{\mathrm{i}}

\newcommand{\rmA}{\mathrm{A}}

\title{Thermal disequilibration of ions and electrons by collisionless plasma turbulence}

\author[a,1]{Yohei~Kawazura}
\author[a,b]{Michael~Barnes}
\author[a,c]{Alexander~A.~Schekochihin}

\affil[a]{Rudolf Peierls Centre for Theoretical Physics, University of Oxford, Clarendon Laboratory, Parks Road, Oxford OX1 3PU, UK}
\affil[b]{Culham Centre for Fusion Energy, Culham Science Centre, Abingdon OX14 3DB, UK}
\affil[c]{Merton College, Oxford OX1 4JD, UK}

\leadauthor{Kawazura} 

\significancestatement{Large-scale astrophysical processes inject energy into turbulent motions and electromagnetic fields, which carry this energy to small scales and eventually thermalise it. How this energy is partitioned between ions and electrons is important both in plasma physics and astrophysics. Here we determine this energy partition via gyrokinetic turbulence simulations and provide a simple prescription for the ion-to-electron heating ratio. We find that turbulence promotes disequilibration of the species: when magnetic-energy density is greater than the thermal energy density, electrons are preferentially heated, whereas when it is smaller, ions are. This is a relatively rare example of nature promoting an ever more out-of-equilibrium state in an environment where particle collisions are not frequent enough to equalise the temperatures of the species.} 

\authorcontributions{A.A.S.\ and M.B.\ designed the research plan. Y.K.\ and M.B.\ developed the simulation code. Y.K.\ conducted the simulations and data analysis. Y.K., A.A.S.\ and M.B.\ devised the physical interpretation of the simulation results and wrote the manuscript.}
\authordeclaration{The authors declare no conflict of interest.}
\correspondingauthor{\textsuperscript{1}To whom correspondence should be addressed. E-mail: yohei.kawazura@physics.ox.ac.uk}


\begin{abstract}
Does overall thermal equilibrium exist between ions and electrons in a weakly collisional, magnetised, turbulent plasma---and, if not, how is thermal energy partitioned between ions and electrons?  This is a fundamental question in plasma physics, the answer to which is also crucial for predicting the properties of far-distant astronomical objects such as accretion disks around black holes. 
In the context of disks, this question was posed nearly two decades ago and has since generated a sizeable literature. 
Here we provide the answer for the case in which energy is injected into the plasma via Alfv\'enic turbulence: collisionless turbulent heating typically acts to disequilibrate the ion and electron temperatures. 
Numerical simulations using a hybrid fluid-gyrokinetic model indicate that the ion-electron heating-rate ratio is an increasing function of the thermal-to-magnetic energy ratio, $\beta_\rmi$: it ranges from $\sim0.05$ at $\beta_\rmi=0.1$ to at least $30$ for $\beta_\rmi \gtrsim 10$. 
This energy partition is approximately insensitive to the ion-to-electron temperature ratio $T_\rmi/T_\rme$. 
Thus, in the absence of other equilibrating mechanisms, a collisionless plasma system heated via Alfv\'enic turbulence will tend towards a nonequilibrium state in which one of the species is significantly hotter than the other, viz., hotter ions at high $\beta_\rmi$, hotter electrons at low $\beta_\rmi$. 
Spectra of electromagnetic fields and the ion distribution function in 5D phase space exhibit an interesting new magnetically dominated regime at high $\beta_i$ and a tendency for the ion heating to be mediated by nonlinear phase mixing (``entropy cascade'') when $\beta_\rmi\lesssim1$ and by linear phase mixing (Landau damping) when $\beta_\rmi\gg1$. 
\end{abstract}

\dates{This manuscript was compiled on \today}
\doi{\url{www.pnas.org/cgi/doi/10.1073/pnas.XXXXXXXXXX}}

\begin{document}

\verticaladjustment{-2pt}

\maketitle
\thispagestyle{firststyle}
\ifthenelse{\boolean{shortarticle}}{\ifthenelse{\boolean{singlecolumn}}{\abscontentformatted}{\abscontent}}{}
\dropcap{I}n many astrophysical plasma systems, such as accretion disks, the intracluster medium and the solar wind, collisions between ions and electrons are extremely infrequent compared to dynamical processes and even to collisions within each species. 
In the effective absence of interspecies collisions, it is an open question whether there is any mechanism for the system to self-organise into a state of equilibrium between the two species and if not, what sets the ion-to-electron temperature ratio. 
This is clearly an interesting plasma-physics question on a fundamental level, but it is also astrophysically important for interpreting observations of plasmas from the heliosphere to the Galaxy, and beyond. 
Historically, the posing of this question 20 years ago in the context of radiatively inefficient accretion flows, and in particular of our own Galactic Centre, Sgr A$^*$ (in which preferential ion heating was invoked to explain low observed luminosity \cite{rees82,narayan95,quataert99}) has prompted a flurry of research and porting of analytical and numerical machinery developed in the context of fusion plasmas and of fundamental turbulence theories to astrophysical problems (see, e.g., \cite{howes06,howes08prl,sch09,howes10,numata10,howes11prl,told15,banon16,kawazura18}, but also \cite{rowan17} and references therein for an alternative strand of investigations). 
In more recent years, heating prescriptions resulting from these investigations have increasingly been in demand for global models aiming to reproduce observations quantitatively (see, e.g., \cite{ressler17,chael18} and references therein). 

In a nonlinear plasma system, turbulence is generally excited by large-scale free-energy sources (e.g., the Keplerian shear flow in a differentially rotating accretion disk), then transferred to ever smaller scales in the position-velocity phase space via a ``turbulent cascade'', and finally converted into thermal energy of plasma particles via microscale dissipation processes. 
This turbulent heating is not necessarily distributed evenly between ions and electrons. 
It may, in principle, lead to either thermal disequilibration or equilibration between ions and electrons, depending on how the ion-to-electron heating ratio changes with the ratio of their temperatures, $T_\rmi/T_\rme$. 
Here we determine this dependence---along with the heating ratio's dependence (which turns out to be much more important) on the other fundamental parameter characterising the thermal state of the plasma, the ratio of the ion-thermal to magnetic energy densities, $\beta_\rmi$. 

This task requires a number of assumptions, many of which are quite simplistic, but are made here to distill what we consider to be the most basic features of the problem at hand. 
We shall assume that the large-scale free-energy injection launches a cascade of perturbations that are anisotropic with respect to the direction of the ambient mean magnetic field and whose characteristic frequencies are Alfv\'enic---we know both from theory \cite{GS95,sch09} and detailed measurements in the solar wind \cite{chen16} that this is what inertial-range turbulence in a magnetised plasma would look like. 
This means that the particles' cyclotron motion can be averaged out at all spatial scales, all the way to the ion Larmor radius and below.
This ``gyrokinetic" (GK) approximation~\cite{frieman82,howes06} leaves out any heating mechanisms associated with cyclotron resonances (because frequencies are low) and with shocks \cite{guo17} (because sonic perturbations are ordered out). 
\add{The amplitude of the fluctuations is assumed to be asymptotically small relative to the mean field, and thus stochastic heating \cite{chandran10} and any other mechanisms relying on finite-amplitude fluctuations~\cite{wu13,gary16,matthaeus16,hughes17,mallet18} are also absent.} 
Furthermore, we assume that ions and electrons individually are near Maxwellian equilibria, but at different temperatures. 
This excludes any heating mechanisms associated with pressure anisotropies \cite{sharma07,sironi15a,kunz18} or significant non-thermal tails in the particle distribution functions \cite{kunz16,chael17}. 
\add{We note that reconnection is allowed within the gyrokinetic model, and so the results obtained here include any heating, ion or electron, that might occur in reconnecting sheets spontaneously formed within the turbulent dynamics.}\footnote{\add{Note, however, that the width of the inertial range that we can afford is necessarily modest. It therefore remains an open question whether reconnecting structures that emerge in collisionless plasma turbulence in extremely wide inertial ranges~\cite{mallet17c,loureiro17b} are capable of altering any of the features of ion-electron energy partition reported here.}}  
\add{Although the GK approximation may be viewed as fairly crude (e.g., it may not always be appropriate to neglect high-frequency fluctuations at ion Larmor scales~\cite{cerri18}), it does a relatively good job of quantitatively reproducing solar wind observations~\cite{howes08prl}; see~\cite{howes08jgr} for a detailed discussion of the applicability of GK model to solar wind. 
In any event, such a simplification is crucial for carrying out multiple kinetic turbulence simulations at reasonable computational cost.}

It can be shown that in GK turbulence, Alfv\'enic and compressive (slow-wave-like) perturbations decouple energetically in the inertial range \cite{sch09}. 
In the solar wind, the compressive perturbations are energetically subdominant in the inertial range \cite{chen16}, although it is not known how generic a situation this is.\footnote{\add{For example, turbulence in accretion flows is mostly driven by the magnetorotational instability (MRI)~\cite{balbus98}. The partition of compressive and Alfv\'enic fluctuations in MRI-driven turbulence is an open question.}} 
At low $\beta_\rmi$, it can be shown rigorously that the energy carried by the compressive cascade will always end up as ion heat. 
Here we shall ignore this heating channel and focus on the Alfv\'enic cascade only, \add{bearing in mind that, at low $\beta_\rmi$, our results likely represent a lower limit on ion heating (another possible source of additional ion heating of low $\beta_\rmi$ is the stochastic heating~\cite{chandran10,mallet18}). }

\section*{Numerical Approach}

An Alfv\'enic turbulent cascade starts in the magnetohydrodynamic (MHD) inertial range, where ions and electrons move in concert. 
Therefore, it is not possible to determine the energy partition between species within the MHD approximation. 
This approximation breaks down and the two species decouple at the ion Larmor scale, $k_\perp \rho_\rmi \sim 1$, where $k_\perp$ is the wave number perpendicular to the mean field. 
At this scale, a certain fraction of the cascading energy is converted into ion heat (via linear and/or nonlinear phase mixing; see below) and the rest continues on as a cascade of ``kinetic Alfv\'en waves'' (KAW), which ultimately heats electrons \cite{sch09}. 
The transition between these two types of turbulence is well illustrated by the characteristic shape of their spectra, familiar from solar-wind measurements at $\beta_\rmi\sim1$ \cite{chen16}: see Fig.~\ref{f:spectra} (middle panel). 

Thus, the energy partition is decided around the ion Larmor scale, where the electron kinetic effects are not important (at least in the asymptotic limit of small electron-to-ion mass ratio). 
We may therefore determine this partition within a hybrid model in which ions are treated gyrokinetically and electrons as an isothermal fluid \cite{sch09}. 
The isothermal electron fluid equations are derived from the electron GK equation via an asymptotic expansion in the electron-to-ion mass ratio $(m_\rme/m_\rmi)^{1/2}$. 
This is valid at scales above the electron Larmor radius and so covers a broad range including both the MHD and ion-kinetic ($k_\perp\rho_\rmi\sim1$) scales. 
In this model, there is an assumed separation of time scales between the fluctuations and the mean fields~\cite{howes06}, which are parametrized by fixed $\beta_\rmi$ and $T_\rmi/T_\rme$ values over the entire course of the simulation.

Our hybrid GK code \cite{kawazura18} (based on \agk~\cite{numata10}, an Eulerian $\delta f$ GK code specialised to slab geometry) substantially reduces the cost of nonlinear simulations. 
It has allowed us to compute the turbulent heating in a proton-electron plasma over a broad parameter range, varying $\beta_\rmi$ from 0.1 to 100 and $T_\rmi/T_\rme$ from 0.05 to 100. 
Most space and astrophysical plasmas have $\beta_\rmi$ and $T_\rmi/T_\rme$ within this range. 
Previous GK simulations of this problem \cite{howes08prl,howes11prl,told15,banon16} were limited to a single point in the parameter space, viz., $(\beta_\rmi, T_\rmi/T_\rme) = (1, 1)$, because of the great numerical cost of resolving both ion and electron kinetic scales.

\add{
In the hybrid code, the phase space of the ion distribution function is spanned by $(x, y, z, \lambda, \varepsilon)$, where $(x, y)$ are the coordinates in the plane perpendicular to the mean magnetic field, $z$ is the coordinate along it, $\lambda = v_\perp^2/v^2$ is the pitch-angle variable, and $\varepsilon = v^2/2$ is the particle kinetic energy. 
The standard resolution used for each simulation was $(n_x, n_y, n_z, n_\lambda, n_\varepsilon) = (64, 64, 32, 32, 16)$. 
In order to verify numerical convergence, we used higher $(x, y)$ resolution $(n_x, n_y, n_z, n_\lambda, n_\varepsilon) = (128, 128, 32, 32, 16)$, higher $z$ resolution $(n_x, n_y, n_z, n_\lambda, n_\varepsilon) = (64, 64, 64, 32, 16)$, and higher $(\lambda, \varepsilon)$ resolution $(n_x, n_y, n_z, n_\lambda, n_\varepsilon) = (64, 64, 32, 64, 32)$ for a few sets of $(\beta_\rmi, T_\rmi/T_\rme)$. 
The range of Fourier modes in the $(x, y)$ plane is set to $0.25 \le k_x\rho_\rmi$,  $k_y\rho_\rmi \le 5.25$ for the standard-resolution runs and $0.125 \le k_x\rho_\rmi$, $k_y\rho_\rmi \le 5.25$ for the high-$(x, y)$-resolution runs. 
In Fig.~\ref{f:QiQe}, we use the highest-resolved simulation available for each point in the parameter space $(\beta_\rmi, T_\rmi/T_\rme)$. 
}

\add{
In order to model the large-scale energy injection, we employ an oscillating Langevin antenna~\cite{tenbarge14}, which excites AWs by driving an external parallel current. 
We set the driven modes to have the oscillation frequency $\omega_a = 0.9\omega_{\rmA0}$, the decorrelation rate $\gamma_a = 0.6\omega_{\rmA0}$, and wave numbers $(k_x/k_{x0},\, k_y/k_{y0},\, k_z/k_{z0}) = (0,\, 1,\, \pm1)$ and $(1,\, 0,\, \pm1)$, where the subscript 0 indicates the smallest wave number in the simulation. 
The antenna amplitude is set to drive critically balanced turbulence, i.e., so as to make the nonlinear cascade rate at the driving scale comparable to the linear wave frequency $\omega_{\rmA0}$. 
}

\add{
The ions have a fully conservative linearised collision operator including pitch-angle scattering and energy diffusion~\cite{abel08,barnes09}. 
The collision frequency is chosen to be $\nu_\rmi = 0.005 \omega_{\rmA0}$, where $\omega_\mr{A0}$ is the AW frequency at the largest scale. 
The ions are thus almost collisionless. 
Since the scale range covered in our simulations is limited, these ``true'' collisions are not sufficient to dissipate all of the energy contained in the ion entropy fluctuations, especially at small spatial scales, where the turbulent eddy-turnover rates are higher. 
Therefore, we use hypercollisions with a collision frequency proportional to $(k_\perp/k_\mr{max})^8$, where $k_\mr{max}$ is the wave number corresponding to the grid scale \cite{howes08prl}. 
While the free energy contained in the perturbed ion distribution function is dissipated by these collisional mechanisms, the physical dissipation mechanisms for the sub-Larmor-scale turbulence destined for electron heating are ordered out by the $(m_\rme/m_\rmi)^{1/2}$ expansion. 
Therefore, we introduce artificial hyperdissipation (hyperviscosity and hyperresistivity) proportional to $(k_\perp/k_\mr{max})^8$ in the isothermal electron fluid equations to terminate KAW cascade~(see \cite{kawazura18} for details). 
We carefully tune the hypercollisionality and hyperdissipation coefficients so as to make the artificial dissipation effective only at the smallest scales.
}
 
\section*{Energy Partition}

\begin{figure*}[t]
  \begin{center}
    \includegraphics*[width=0.75\textwidth]{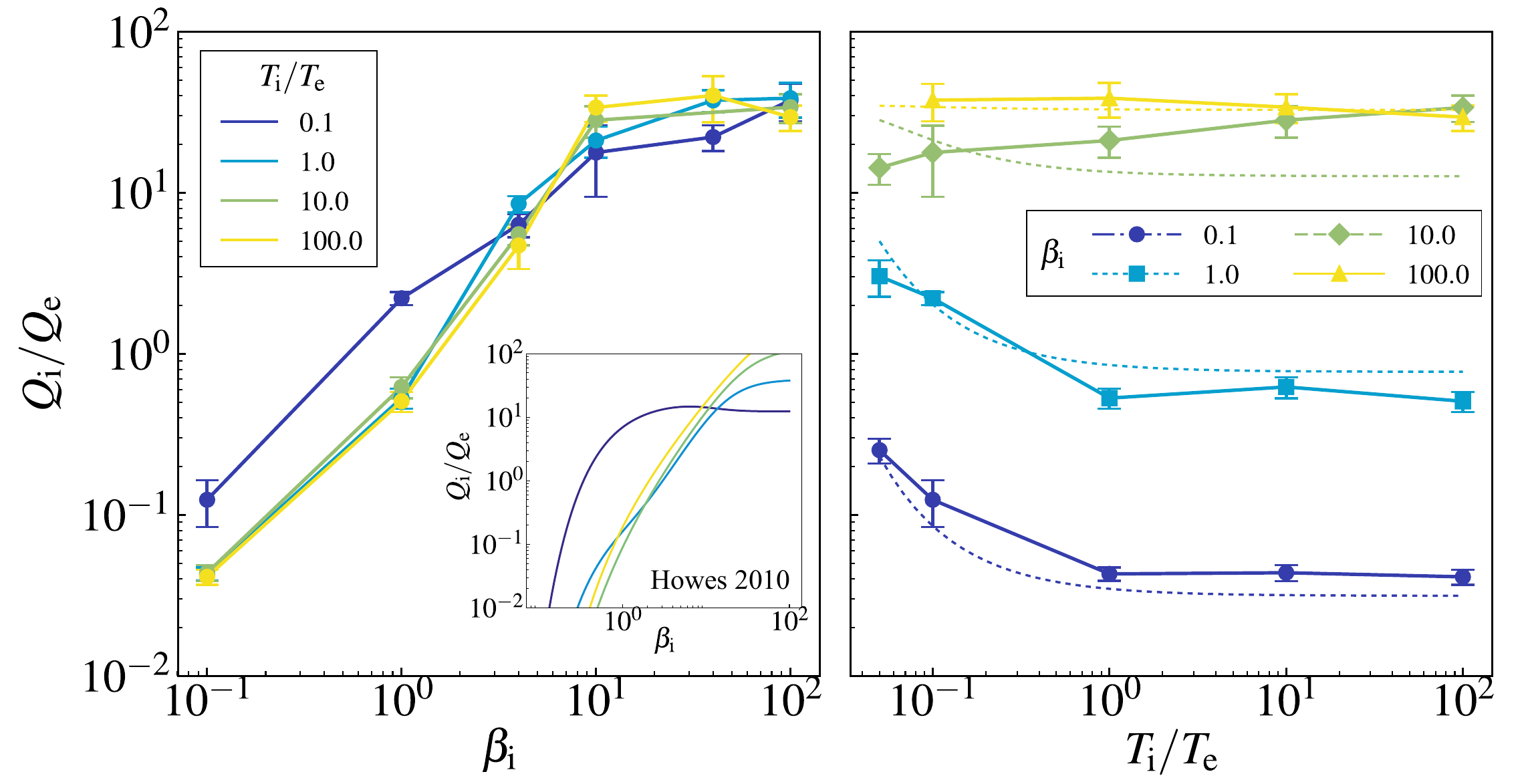}
  \end{center}
  \caption{
    The ion-to-electron heating ratio $Q_\rmi/Q_\rme$ vs.\ $\beta_\rmi$ (left) and $T_\rmi/T_\rme$ (right). 
    We take the time average in the steady state for a period $\gtrsim 5 t_\rmA$, where $t_\rmA$ is Alfv\'en time at the box scale. 
    The error bars show the standard deviation of the time series. 
    The dotted lines in the right panel show the fitting formula (\ref{eq:fit}). 
    The inset in the left panel is $Q_\rmi/Q_\rme$ vs. $\beta_\rmi$ calculated via the model proposed in~\cite{howes10}, based on linear theory: note the much lower ion heating at low $\beta_i$, absence of a ``ceiling'' at high $\beta_\rmi$  and a more dramatic deviation of the case of cold ions (low $T_\rmi/T_\rme$) from the general trend.
  }
  \label{f:QiQe}
\end{figure*}

\begin{figure*}[t]
  \begin{center}
    \includegraphics*[width=1.0\textwidth]{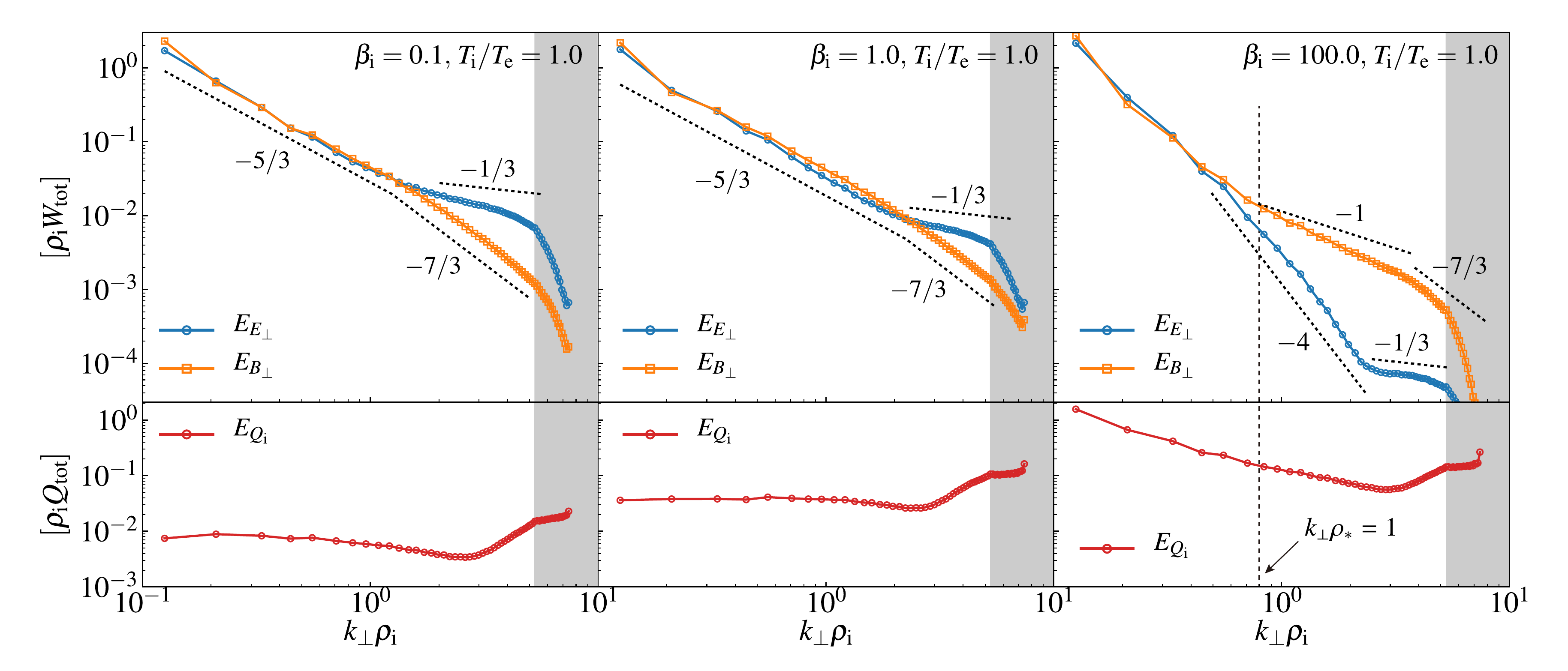}
  \end{center}
  \caption{
    Spectra of magnetic (blue) and electric (orange) perturbations, in units of total free energy ($W_{\rm tot}$) times $\rho_\rmi$, for three representative values of $\beta_\rmi=0.1, 1, 100$ and $T_\rmi/T_\rme=1$. 
    The shaded region shows the corner modes in $(k_x, k_y)$ plane\add{, where the $(x,y)$ plane is perpendicular to the ambient magnetic field direction $z$.} 
    Various theoretical slopes are shown for reference: $k_\perp^{-5/3}$ in the inertial range (standard MHD turbulence \cite{GS95}); $k_\perp^{-7/3}$ for magnetic and $k_\perp^{-1/3}$ for electric fields in the sub-ion-Larmor range (KAW cascade \cite{cho04,sch09}); and $k_\perp^{-1}$ for the purely magnetic cascade at high $\beta_\rmi$ (similar to subviscous MHD cascade \cite{cho02visc}; the scale $\rho_*$ at which this starts, defined in the text is also shown). 
    Clearly, at these resolutions, a definitive determination of spectral slopes is not feasible. 
    Lower panels show ion heating rate vs.~$k_\perp$, in units of total injected power ($Q_{\rm tot}=Q_\rmi + Q_\rme$) times $\rho_\rmi$. 
    The uptick in ion heating at the smallest scales is due to ion hyperresistivity and hyperviscosity.
  \add{We note that halving the box size for the $\beta_\rmi=100$ simulation results in only a 10\% change to $Q_\rmi/Q_\rme$ (which is smaller than the error due to finite-time averaging), suggesting that this result is independent of injection scale.}
  }
  \label{f:spectra}
\end{figure*}

The main result of our simulations is given in Fig.~\ref{f:QiQe}, which shows the dependence of the ratio of the time-averaged ion and electron heating rates $Q_\rmi/Q_\rme$ on $\beta_\rmi$ and $T_\rmi/T_\rme$. 
The left panel shows that $Q_\rmi/Q_\rme$ increases as $\beta_\rmi$ increases regardless of $T_\rmi/T_\rme$. 
When $(\beta_\rmi, T_\rmi/T_\rme) = (1, 1)$, we find $Q_\rmi/Q_\rme \approx 0.6$, in good agreement with the result found in the full GK simulation studies that resolved the entire range from MHD to electron kinetic scales~\cite{told15,banon16}. 
We find that ions receive more energy than electrons when $\beta_\rmi \gtrsim 1$ while electron heating is dominant in the low-$\beta_\rmi$ regime. 

\subsection*{Low Beta}

In the limit $\beta_\rmi\to0$, our results suggest $Q_\rmi/Q_\rme\to 0$, which is physically intuitive: in this regime, the ion thermal speed is much smaller than the Alfv\'en speed, so ions cannot interact with Alfv\'enic perturbations and so the cascade of the latter smoothly turns into a sub-Larmor KAW cascade, without any energy being diverted into ions \cite{zocco11}. 
This ``smooth'' transition is manifest when one examines the energy spectra in this regime (Fig.~\ref{f:spectra}, left panel). 

\add{
The scale where the ion heating occurs is apparent in the bottom panels of Fig.~\ref{f:spectra}. 
For low to moderate $\beta_\rmi$, the ion heating is dominated by grid-scale hyperdissipation. 
This is consistent with the previous full gyrokinetic simulation with $\beta_\rmi = 1$~\cite{howes11prl,told15,banon16}, where the ion heating peaked at $20 \lesssim k_\perp\rho_\rmi \lesssim 30$.
In contrast, the ion heating for high $\beta_\rmi$ occurs predominantly at large-scales, which is revealed for the first time in this study (see below).
}

\subsection*{High Beta}

In the opposite limit of high $\beta_\rmi$, simulations show that $Q_\rmi/Q_\rme$ increases and appears to tend to a constant $\simeq 30$ for $\beta_\rmi \gtrsim 10$. 

The physics behind this result is more complicated. 
In a high-$\beta_\rmi$ plasma, Alfv\'en waves (AW) are damped at a rate that peaks around $k_\perp\rho_\rmi \sim \beta_\rmi^{-1/4}$, where it is comparable to their propagation frequency: namely, in the limit $\beta_\rmi\gg1$, 
the complex frequency is \cite{howes06,kunz18}
\begin{equation}
\omega = |k_\para| v_\rmA\left[\pm\sqrt{1-(k_\perp\rho_*)^4} - i(k_\perp\rho_*)^2\right],  
\end{equation}
where $\rho_* = (3/4\pi^{1/4}\sqrt{2})\beta_\rmi^{1/4}\rho_\rmi$. 
At $k_\perp\rho_* > 1$, AW can no longer propagate and at $k_\perp\rho_* \gg 1$, damping peters out for magnetic perturbations ($\omega\approx-i|k_\para| v_\rmA/2k_\perp^2\rho_*^2$), but becomes increasingly strong for velocity (electric-field) perturbations ($\omega\approx-i|k_\para| v_\rmA2k_\perp^2\rho_*^2$). 
The situation resembles an overdamped oscillator, with magnetic field in the role of displacement. 
This means that at $k_\perp\rho_*\sim1$, the MHD Alfv\'enic cascade is partially damped and partially channelled into a purely magnetic cascade, as is indeed evident in the right panel of Fig.~\ref{f:spectra} (this resembles the subviscous cascade in high-magnetic-Prandtl-number MHD and, similarly to it \cite{cho02visc}, might be exhibiting a $k_\perp^{-1}$ spectrum, arising from nonlocal advection of magnetic energy by $\rho_*$-scale motions). 
The magnetic cascade extends some way into the sub-ion-Larmor range, but eventually, at $k_\perp\rho_\rmi\gg1$, it must turn into a KAW cascade. 
While the sort of spectra that we find at $\beta_\rmi\lesssim 1$ (Fig.~\ref{f:spectra}, left and middle panels) are very similar to what has been observed both in numerical simulations \cite{howes08prl,howes11prl,told15,groselj18prl,cerri18,franci18} and in solar-wind observations \cite{chen16} at $\beta_\rmi\sim1$, the high-$\beta_\rmi$ spectra described above have not been seen before and represent an interesting new type of kinetic turbulence.

Thus, there is a finite wave-number interval of strong damping around $k_\perp\rho_*\sim1$. 
In a ``critically balanced'' turbulence, $|k_\para|v_\rmA$ is of the same order as the cascade rate, so this damping will divert a finite fraction of total cascaded energy into ion heat (this is manifest in Fig.~\ref{f:vspectrum}, panel D). 
Exactly what fraction it will be is what our numerical study tells us. 
We do not have a quantitative theory that would explain why $Q_\rmi/Q_\rme$ should saturate at the value that we observe numerically (which, based on a resolution study, appears to be converged). 
Presumably, this is decided by the details of the operation of ion Landau damping in a turbulent environment (a tricky subject \cite{sch16,adkins18,meyrand18}) and by the efficiency with which energy can be channelled from the MHD scales into the magnetic cascade below $\rho_*$ and the KAW cascade below $\rho_\rmi$. 
In the absence of a definitive theory, $Q_\rmi/Q_\rme\approx 30$ should be viewed as an ``experimental'' result.

\subsection*{Relation to Standard Model Based on Linear Damping}

It is instructive to compare $Q_\rmi/Q_\rme$ obtained in our simulations with the simple theoretical model for the turbulent heating proposed in \cite{howes10}, which has been used as a popular prescription in global disk models \cite{ressler17,chael18}. 
The model is based on assuming (i) continuity of the magnetic-energy spectrum across the ion-Larmor-scale transition, (ii) linear Landau damping as the rate of free-energy dissipation leading to ion heating and (iii) critical balance between linear propagation and nonlinear decorrelation rates. 
As evident in the inset in Fig.~\ref{f:QiQe}, the model gives a broadly correct qualitative trend, but produces some noticeable quantitative discrepancies: notably, much lower ion heating at low $\beta_\rmi$ and an absence of the ``ceiling'' on $Q_\rmi/Q_\rme$ at high $\beta_\rmi$. 

This is perhaps not surprising, for a number of reasons. 
First, the Landau damping rate is not, in general, a quantitatively good predictor of the rate at which linear phase mixing would dissipate free energy in a driven system \cite{kanekar15}. 
Indeed, we have found that an approximation such as $E_{Q_\rmi}(k_\perp)\propto\mathrm{Im}\,\omega(k_\para,k_\perp)E_{B_\perp}(k_\perp)$ (with $\omega$ the linear frequency and $k_\para$ either directly measured or inferred from the critical-balance conjecture) did not reproduce quantitatively the heating spectra shown in the lower panels of Fig.~\ref{f:spectra}. 
Secondly, at high $\beta_\rmi$, the model of \cite{howes10} does not treat turbulence in the no-propagation region at $k_\perp\rho_*$ as a nonlocally driven magnetic cascade, choosing rather to smooth the frequency gap between the Alfv\'en waves and KAW. 
Thirdly, at low $\beta_\rmi$, as we are about to see below, the ion heating is controlled by the nonlinear, rather than linear, phase mixing (``entropy cascade'' \cite{sch09,tatsuno09,plunk10,cerri18}). 

\subsection*{Temperature Disequilibration}

Apart from the $\beta_\rmi$ dependence, the key finding of our simulations is that $Q_\rmi/Q_\rme$ is mostly insensitive to $T_\rmi/T_\rme$ (keeping $\beta_\rmi$ constant; see the right panel of Fig.~\ref{f:QiQe}). 
Some dependence on $T_\rmi/T_\rme$ does exist when $\beta_\rmi\lesssim1$ and $T_\rmi/T_\rme$ is small (for $\beta_\rmi\ll1$, this is the ``Hall limit'' of GK \cite{sch09}). 
This dependence is redistributive: colder ions are heated a little more. 
At low $\beta_{\rmi}$, most of the energy still goes into electrons, but at $\beta_{\rmi}\sim1$, the effect might be of some help in restoring some parity between ions and electrons because $Q_\rmi/Q_\rme>1$ at low $T_\rmi/T_\rme$ and $Q_\rmi/Q_\rme<1$ at high $T_\rmi/T_\rme$. 

Overall, we see that whether ions and electrons are already disequilibrated or not makes relatively little difference to the heating rates---there is no intrinsic tendency in the collisionless system to push the two species towards equilibrium with each other (except at $\beta_\rmi\sim1$). 
In fact, in the absence of ion cooling and at constant magnetic field, turbulent heating would gradually increase beta and thus push the system towards a state of dominant ion heating and hence hotter ions.
Runaway increase of $T_\rmi/T_\rme$ can be envisioned if $T_\rme$ is capped by, e.g., radiative cooling. 

\subsection*{Fitting Formula}

For a researcher who is interested in using these results in global models (as in, e.g., \cite{ressler17,chael18}), here is a remarkably simple fitting formula, which, without aspiring to ultra-high precision, works quite well over the parameter range that we have investigated (see Fig.~\ref{f:QiQe}, right panel): 
\begin{equation}
\label{eq:fit}
\frac{Q_\rmi}{Q_\rme} = \frac{35}{1 + (\beta_\rmi/15)^{-1.4} e^{-0.1\, T_\rme/T_\rmi}}. 
\end{equation}

\section*{Phase-Space Cascades}

\begin{figure}[t]
  \centerline{\includegraphics[height=0.5\textheight]{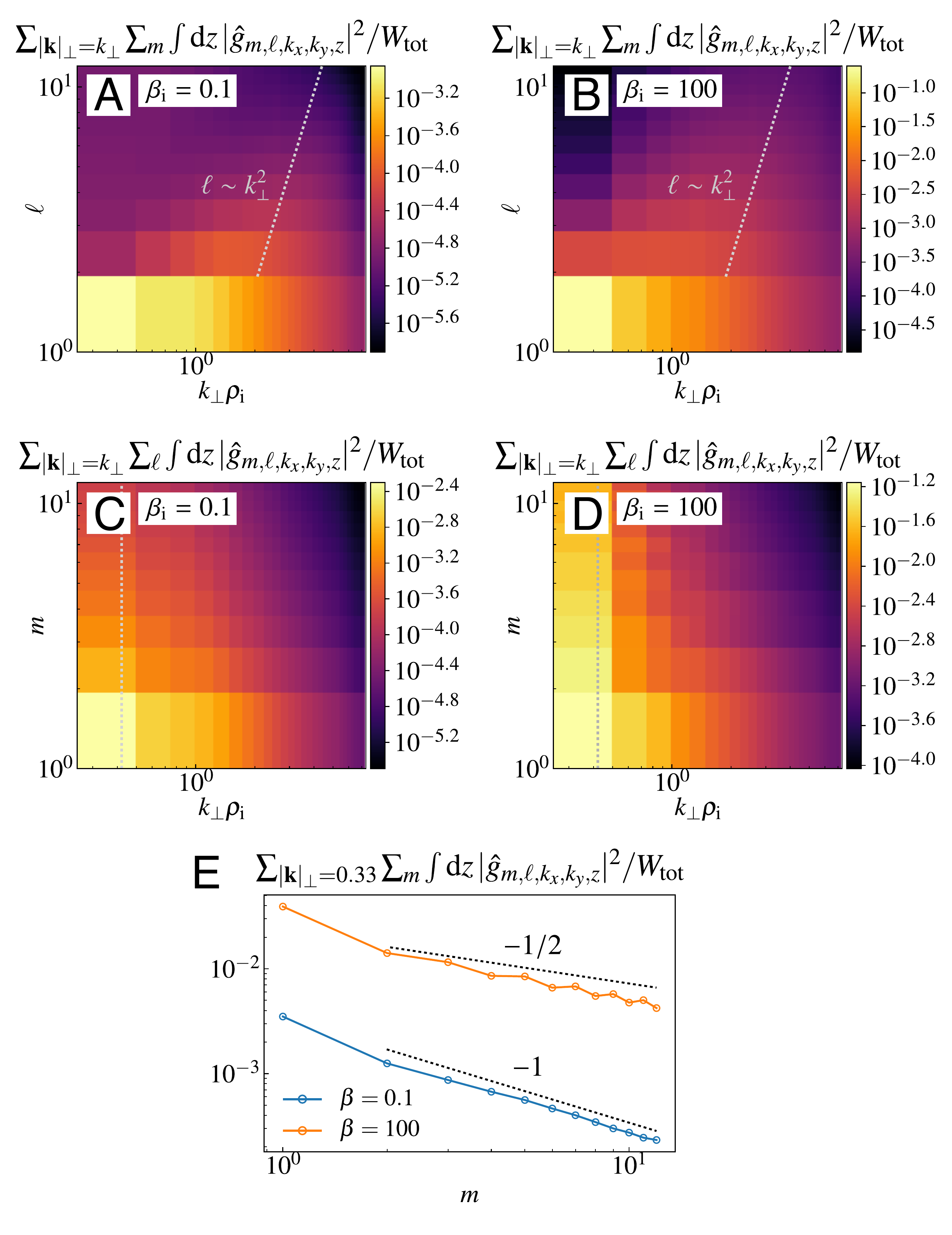}}
  \caption{
    Phase-space spectra of the gyroaveraged perturbed ion distribution function: $|\hat{g}|^2$ in Fourier--Laguerre space $(k_\perp,\ell)$ (panels A, B) and Fourier--Hermite space $(k_\perp,m)$ (panels C, D) for $T_\rmi/T_\rme=1$, $\beta_\rmi = 0.1$ (panels A, C) and $\beta_\rmi = 100$ (panels B, D). 
    Panel E: Hermite spectrum at $\krhoi = 0.33$, i.e., a cut along the dotted line in panels C and D, for $\beta_\rmi = 0.1$ (blue) and 100 (orange). 
    Note the standard $m^{-1/2}$ slope associated with linear phase mixing \cite{zocco11,kanekar15} at high $\beta_\rmi$ and a steeper $m^{-1}$ slope at lower $\beta_\rmi$, indicating suppressed phase mixing (cf.~\cite{adkins18,meyrand18}).}
  \label{f:vspectrum}
\end{figure}

One of the more fascinating developments prompted by the interest in energy partition in plasma turbulence has been the realisation that, in a kinetic system, we are dealing with a free-energy cascade through the entire phase space, with energy travelling from large to small scales in both position and velocity space \cite{sch09,tatsuno09,plunk10,hatch14,sch16,parker16,servidio17,adkins18,cerri18,eyink18,pezzi18}. 
This is inevitable because the plasma collision operator is a diffusion operator in phase space and so the only way for a kinetic system to have a finite rate of dissipation at very low collisionality is to generate small phase-space scales---just like a hydrodynamic system with low viscosity achieves finite viscous dissipation by generating large flow-velocity gradients. 
The study of velocity-space cascades in kinetic systems is still in its infancy---but advances in instrumentation and computing mean that the amount of available information on such cascades in both real (space-) physical plasmas \cite{servidio17} and their numerical counterparts \cite{cerri18,pezzi18,meyrand18} is rapidly increasing. 
Let us then investigate the nature of the phase-space cascade in our ion-heating simulations. 

In low-frequency (GK) turbulence, there are two routes for the velocity-space cascade: linear phase mixing, also known as Landau damping \cite{landau46}, produces small scales in the distribution of the velocities parallel to the magnetic field ($v_\para$) \cite{hammett92,kanekar15}, whereas the cascade in the perpendicular velocities ($v_\perp$) is brought about by nonlinear phase mixing, or ``entropy cascade'', associated with particles following Larmor orbits (whose radii are $\propto v_\perp$) sampling spatially decorrelated electromagnetic perturbations \cite{sch09,tatsuno09,plunk10}. 
The latter mechanism switches on at spatial scales for which the Larmor radius is finite, i.e., at $k_\perp\rho_\rmi\gtrsim 1$. 
While these velocity-space cascades are interesting in themselves as fundamental phenomena setting the structure of plasma turbulence in phase space, they also give us a handle on whether the ion heating tends to be parallel or perpendicular (this could become important if we asked, e.g., towards what kind of pressure-anisotropic states turbulence pushes the plasma). 

We employ the Hermite--Laguerre spectral decomposition of the gyroaveraged perturbed distribution function $g = \langle\delta f\rangle$~\cite{mandell18}: 
\begin{equation}
  \hat{g}_{m,\ell} = \int_{-\infty}^{\infty}\!\rmd v_\para\f{H_m(v_\para/v_\mr{thi})}{\sqrt{2^m m!}}
\int_{0}^{\infty}\!\rmd (v_\perp^2) L_\ell(v_\perp^2/v_\mr{thi}^2)g(v_\para, v_\perp^2),
\label{eq:g_def}
\end{equation}
where $H_m(x)$ and $L_\ell(x)$ are the Hermite and Laguerre polynomials. 
In this language, higher $m$ and $\ell$ correspond to smaller scales in $v_\para$ and $v_\perp$, respectively. 
Fig.~\ref{f:vspectrum} shows the phase-space spectra of the ion entropy ($|\hat{g}|^2$, the contribution of the perturbed ion distribution function to the free energy \cite{sch09}) for $\beta_\rmi = 0.1$ and $\beta_\rmi = 100$ cases with $T_\rmi/T_\rme=1$. 
We see that the distribution of the free energy and, consequently, the nature of its cascade through phase space changes with $\beta_\rmi$. 

\subsection*{Low Beta}

At low $\beta_\rmi$, linear phase mixing is suppressed (see panel C; this is because ions' thermal motion is slow compared to the phase speed of the Alfv\'enic perturbations), so most of the ion entropy is cascaded simultaneously to large $k_\perp\rho_\rmi$ and $\ell$ by nonlinear phase mixing (panel A) before being thermalised by collisions, giving rise to (perpendicular) ion heating. 
The Fourier--Laguerre spectrum contains little energy at high $\ell$ when $k_\perp\rho_\rmi<1$ (because plasma dynamics are essentially drift-kinetic at these scales and there is no phase mixing in $v_\perp$), but at $k_\perp\rho_\rmi>1$ it is consistent with aligning along $\ell\sim (k_\perp\rho_\rmi)^2$. 
This is a manifestation of the basic relationship between the velocity and spatial scales, $\delta v_\perp/v_\mr{thi} \sim 1/k_\perp\rho_\rmi$, that is characteristic of sub-Larmor entropy cascade \cite{sch09,tatsuno09,plunk10} ($\delta v_\perp/v_\mr{thi} \sim 1/\sqrt{\ell}$ follows from the trigonometric asymptotic of Laguerre polynomials at high $\ell$). 
Similar ``diagonal'' structure has previously been found in 4D electrostatic GK simulations~\cite{tatsuno10} and in 6D electromagnetic hybrid-Vlasov simulations~\cite{cerri18}. 
Note also that for the case $(\beta_\rmi, T_\rmi/T_\rme) = (1, 1)$, Ref.~\cite{banon16} compared the contributions to ion heating from the $v_\perp$ and $v_\para$ parts of the collision operator and also concluded that the nonlinear phase mixing was the dominant process. 

\subsection*{High Beta}

In contrast, at high $\beta_\rmi$, most ion entropy is channelled to high $m$ at $k_\perp\rho_\rmi<1$ (panel D) by linear phase mixing, as is indeed confirmed by the characteristic $m^{-1/2}$ slope of the Hermite spectrum \cite{zocco11,kanekar15} (see panel E; at low $\beta_\rmi$, the Hermite spectrum is steeper, implying very little dissipation \cite{sch16,adkins18}). 
These perturbations are then thermalised at high $m$ by collisions. 
Thus, the preferential heating of ions at high $\beta_\rmi$ is parallel and occurs via ordinary Landau damping.\footnote{We make this statement with some caution. The velocity resolution of our simulations is necessarily limited, so our plasma has a certain effective collisional cutoff $m_\rmc$ (typically, $m_\rmc\sim10$). 
The order of limits $m_\rmc\to\infty$ and $\beta_\rmi\to\infty$ may matter to the system's ability to block linear phase-mixing via the stochastic echo effect because the rate at which free energy is transferred from $m$ to $m+1$ by linear phase mixing is $\sim |k_\para|v_\mr{thi}/\sqrt{m}$ \cite{sch16,adkins18} whereas the nonlinear advection rate in a critically balanced Alfv\'enic turbulence is $\sim |k_\para|v_\rmA = |k_\para|v_\mr{thi}/\sqrt{\beta_\rmi}$. 
At the highest values of $\beta_i$, our simulations have $m_\rmc < \beta_\rmi$, so the effective collisionality may interfere with the echo. 
If, at infinite resolution (i.e., in an even less collisional plasma than we simulate currently), the echo is restored, ion heating at $\beta_\rmi\gg m_\rmc$ may be all via the entropy cascade.}  

\section*{Discussion}

To discuss an example of astrophysical consequences of our findings, let us return briefly to the curious case of low-luminosity accretion flows---most famously, the supermassive black hole Sgr A$^*$ at our Galaxy's centre. 
The theories that have been advanced to explain the observed low luminosity range between two basic scenarios: in the first, $Q_\rmi/Q_\rme\gg1$ and so most of the thermal energy is deposited into non-radiating ions, which are swallowed by the black hole \cite{rees82,narayan95,quataert99}; in the second, $Q_\rmi/Q_\rme\sim 1$ but the accretion rate is very small, with most of the plasma being carried away by outflows \cite{blandford99}. 
Determining which of these is closer to the truth is tantamount to identifying the fate of the accreting matter. 
\add{
The low accretion rate scenario has gradually become more widely accepted~\cite{yuan03,sharma07,xie12}, whereas early studies used the high-$Q_\rmi/Q_\rme$ scenario~\cite{narayan95,esin97}.
The value $Q_\rmi/Q_\rme \simeq 30$ that we have found for moderately high values of $\beta_\rmi$ is about ten times larger than the value used today. 
However, even with this value, the accretion rate must be much smaller than the Bondi rate (see Fig.~1 in Ref.~\cite{xie12}), given the observational fact that the outflow is present~\cite{yuan03,wong11}.
}
Within this scenario, the relative amount of electron heating in the low-$\beta_\rmi$, central region of the disk turns out to be crucial to enable a detectable jet: Ref.~\cite{chael18} found a radiating jet in global simulations using the linear prescription with very low ion heating \cite{howes10} and no visible jet with a more equitable heating model \cite{rowan17}. 
Our heating prescription is perhaps closer to \cite{howes10} in that regard, but not as extreme---it would be interesting to see what effect this has on global models of accreting systems. 

On a broader and perhaps more fundamental level, we have shown that turbulence is capable of pushing weakly collisional plasma systems away from inter-species thermal equilibrium---depending on whether $\beta_\rmi$ is high or low, it favours preferential thermalisation of turbulent energy into ions or electrons, respectively (although at $\beta_\rmi\sim1$, there is some tendency to restoration of species equality). 
This is a relatively rare example of turbulence failing to promote Le Chatelier's principle and instead causing a disequilibration of a collisionless system. 




\acknow{
We thank S.~Balbus, B.~Chandran, S.~Cowley, W.~Dorland, C.~Gammie, G.~Howes, M.~Kunz, N.~Loureiro, A.~Mallet, R.~Meyrand, F.~Parra and E.~Quataert for fruitful discussions and suggestions. 
This work was supported by the STFC grant ST/N000919/1. 
AAS was also supported in part by EPSRC grant EP/M022331/1. 
For the simulations reported here, the authors acknowledge the use of ARCHER through the Plasma HEC Consortium EPSRC grant number EP/L000237/1 under projects e281-gs2, the EUROfusion HPC (Marconi--Fusion) under project MULTEI, the Cirrus UK National Tier-2 HPC Service at EPCC funded by the University of Edinburgh and EPSRC (EP/P020267/1) and the University of Oxford's ARC facility.
}

\showacknow 


\bibliography{bib_JPP.bib}
\end{document}